\documentclass[aps,prb,twocolumn,superscriptaddress]{revtex4}
\usepackage{graphicx}
\usepackage{amssymb}
\usepackage{amsmath}
\usepackage{graphicx}
\usepackage{epsfig}
\usepackage{color}
\usepackage{ulem}
\usepackage{hyperref}
\usepackage{appendix}
\newcommand{\dd}{\mathrm{d}}
\begin{document}
\title{Learning disentangled representation for classical models}
\author{Dongchen Huang}
\author{Danqing Hu}
\affiliation{Beijing National Laboratory for Condensed Matter Physics and Institute of Physics, Chinese Academy of Sciences, Beijing 100190, China}
\affiliation{University of Chinese Academy of Sciences, Beijing 100049, China}
\author{Yi-feng Yang}
\email[]{yifeng@iphy.ac.cn}
\affiliation{Beijing National Laboratory for Condensed Matter Physics and Institute of Physics, Chinese Academy of Sciences, Beijing 100190, China}
\affiliation{University of Chinese Academy of Sciences, Beijing 100049, China}
\affiliation{Songshan Lake Materials Laboratory, Dongguan, Guangdong 523808, China}

    \begin{abstract}
Finding disentangled representation plays a predominant role in the success of modern deep learning applications, but the results lack a straightforward explanation. Here we apply the information bottleneck method and its $\beta$-VAE implementation to find the disentangled low-dimensional representation of classical models. For the Ising model, our results reveal a deep connection between the disentangled features and the physical order parameters, and the widely-used Bernoulli decoder is found to be learning a mean-field Hamiltonian at fixed temperature. This analogy motivates us to extend the application of $\beta$-VAE to more complex classical models with non-binary variables using different decoder neural network and propose a modified architecture $\beta^2$-VAE to enforce thermal fluctuations in generated samples. Our work provides a way to design novel physics-informed algorithm that can yield learned features in potential correspondence with real physical properties.
    \end{abstract}
    
    \maketitle
    
\section{Introduction}
Representation learning lies in the core of modern deep learning. With given tasks such as in supervised learning, deep learning methods have achieved unprecedented success in computer vision \cite{7780459,NIPS2012_4824,10.1007/978-3-319-46493-0_38} and deep reinforcement learning \cite{mnih2013playing,Schrittwieser2019MasteringAG,AlphaStar}. However,  it is still very challenging to establish clear criteria for the learned representations \cite{10.1007/978-3-319-46493-0_38, Locatello19}. In practice, algorithmic designs are guided by some \emph{meta-priors} on general premises about the relevant domain knowledge \cite{6472238}. One of the most popular and typical meta-priors is \emph{disentanglement}, which in a recent line of works \cite{6472238,pmlr-v80-kim18b,DLYGG,betaVAE} has been argued to be a good characteristic of representations. A key intuition of disentanglement is that a good representation requires the data to have some independent generative factors that are explainable and semantically meaningful and have the ability of reconstruction, which is also the goal of generative modeling. Two typical methods are often used for the generative model. One is the likelihood-free method which parameterizes the probability distribution directly with deep neural network \cite{goodfellow2014generative,chen16}. The other is the likelihood-based method \cite{Kingma13,Rezende14,pmlr-v80-kim18b,betaVAE}, which assumes an explicit probability distribution of the input data.  

The most prevalent likelihood-based methods are restrictive Boltzmann machines (RBMs) and variational autoencoders (VAEs). RBMs \cite{Smolensky1986} and their quantum extensions \cite{Amin2016QuantumBM} have many applications in the field of many-body physics \cite{Koch-Janusz2018,Nagy2019,Hartmann2019,Vicentini2019,Yoshioka2019} and quantum information \cite{Torlai2016,Torlai2018,Torlai2018b}, such as approximating wave functions \cite{Carleo2017}, performing quantum state tomography \cite{Torlai2018}, and learning the Ising model even near criticality \cite{Alan2018}. By contrast, VAEs are relatively less applied in physics, although they have been extensively explored in machine learning. VAEs have the advantage of learning the latent representation and the probability distribution of the input data simultaneously \cite{Rezende14,Kingma13}, which is lacking in RBMs due to the difficulty in interpreting the latent representations and generating high-fidelity images. VAEs can also well capture the physical properties with the usage of convolutional layers \cite{DAngelo2020LearningTI} and make further predictions \cite{Efthymiou2019}. Their linear versions, namely the principal component analysis (PCA) and its kernel extension \cite{scholkopf1997}, have been shown to be a useful extractor to identify phase transitions \cite{Wang2016}. It is therefore interesting to explore more applications of VAEs in physics.

On the other hand, combining prior knowledge into learning algorithms also plays a crucial role in designing neural network \cite{Li2018,Iso2018,Kim2018,Gao2018AQM,Lloyd2018,Zoufal2019}. In physics, people also look for simplified representation of the complex world. The interaction of physics and representation learning may lead to potential improvement of both fields.  The exploration of VAEs in physics may potentially improve the expressive power and interpretability. In fact, previous machine learning applications have mostly focused on studying \emph{raw} data rather than controlling latent representations. It is natural to ask what is the disentangled representation for the \emph{physical} data as reflected in the latent space, and to what extent the learned representation resembles the true physical knowledge. 

To simplify the discussions, here we make the attempt to address these questions by studying the classical Ising model as an example. We apply the state-of-the-art disentanglement learning method $\beta$-VAE to approximate the \emph{sufficient statistic} of configuration samples of the Ising model and show that introducing an extra hyperparameter can simultaneously improve the quality of both representation and reconstruction. This leads to better interpretability of the disentangled representation and projects different physical properties onto different axes of the latent space, and the distance in input space is protected with a well-trained encoder via maximum mean discrepancy \cite{gretton12a}. We further find that the popular Bernoulli decoder can be viewed as a mean-field solver for an effective local Hamiltonian, providing a physical interpretation of the last layer of the decoder. This motivates us to propose novel physics-informed algorithm that can yield learned features in potential correspondence with real physical properties. We then apply our method to more complex models with non-binary classical variables and confirm its general applicability.

\section{Method}
We first introduce the $\beta$-VAE method recently developed based on the information bottleneck principle. The latter provides a candidate criteria for good representations and has the advantage to be approximately implemented within the framework of generative modelling.

\subsection{Information bottleneck}
A good representation of the data should satisfy the conditions of sufficiency and simplicity. Sufficiency means that the representation should well capture key properties of the data distribution. The condition of sufficiency can be realized by introducing the so-called \emph{sufficient statistic}. For a given random variable $X$ with the probability distribution $P(X)$, its sufficient statistic is a random variable $Z$ that contains all the information of $X$ in the sense that $I(Z; X)=H(X)$, where $H(X)$ is the Shannon entropy of $X$ and $I(Z; X)$ is the \emph{mutual information} between $Z$ and $X$ defined by
\begin{equation}
I(Z; X) = \sum_{Z, X}P(Z, X) \ln \frac{P(Z, X)}{P(Z)P(X)}.
\end{equation}
Here $P(Z, X)$ is the joint probability of $Z$ and $X$. We have $I(Z; X) = 0$ if and only if $X$ and $Z$ are independent. Thus the mutual information illustrates the correlation between a pair of random variables. Here, it is used to characterize the sufficiency of $Z$ in representing $X$.

Simplicity means that the representation should be as simple as possible. A common empirical assumption is disentanglement, where high dimensional data can be described by a hidden low dimensional set of independent factors (disentanglement). To achieve this, it was shown recently that one should also introduce explicit inductive biases referring to certain task $Y$ \cite{Locatello19}, so that the optimal representation $Z$ may ignore some irrelevant information of $X$ but encode the most relevant information of $X$ with respect to $Y$.

Obviously, the choice of representation $Z$ is not unique. Since for any transformation $S(X)$, the mutual information satisfies the data processing inequality $I(S(X);X)\le H(X)$, the information bottleneck principle suggests that a good representation may be found by solving the following constrained optimization problem \cite{Tishby99,Shwartz-Ziv17}:
\begin{equation}
Z = \mathop{\mathrm{argmax}}_{S(X);\atop I(S(X); X)\leq I_c} I(S(X); Y),
\label{Eq:sufficient statistic}
\end{equation}
where $Y$ is the given task (e.g. classification), $I_c$ controls how much information is to be ignored, and $S(X)$ is a transformation of the random variable $X$. If the mutual information of $Z$ and $X$ equals the Shannon entropy of $X$, the optimal representation is also sufficient statistic by definition. For Gaussian distribution of $X$, the sufficient statistic  is simple and well-known, \emph{i.e.}, its mean and variance.

But in general, it is an intractable task to obtain the exact formulation of optimal $S(X)$. We can further approximate the above optimization problem by computing
\begin{equation}
Z = \mathop{\mathrm{argmax}}_{Z} \left[I(Z; Y)-\beta I(Z; X)\right],
\label{Eq:IB}
\end{equation}
where the Lagrange multiplier $\beta$ controls the relevance of $Z$ with respect to $X$. This links closely to the generative modelling where the task $Y$ is nothing but the intrinsic probability distribution of the data and the random variable $X$ obeys the empirical distribution of the samples. Thus, in the following, we will not distinguish the random variables $X$ and $Y$. The meta-prior of disentanglement appears in the assumption of the structure of the latent variable $Z$, where high dimensional data can be described by a hidden low dimensional set of independent parameters.  

\subsection{$\beta$-VAE}

\begin{figure}[b]
  \centering
  \includegraphics[width=0.48\textwidth]{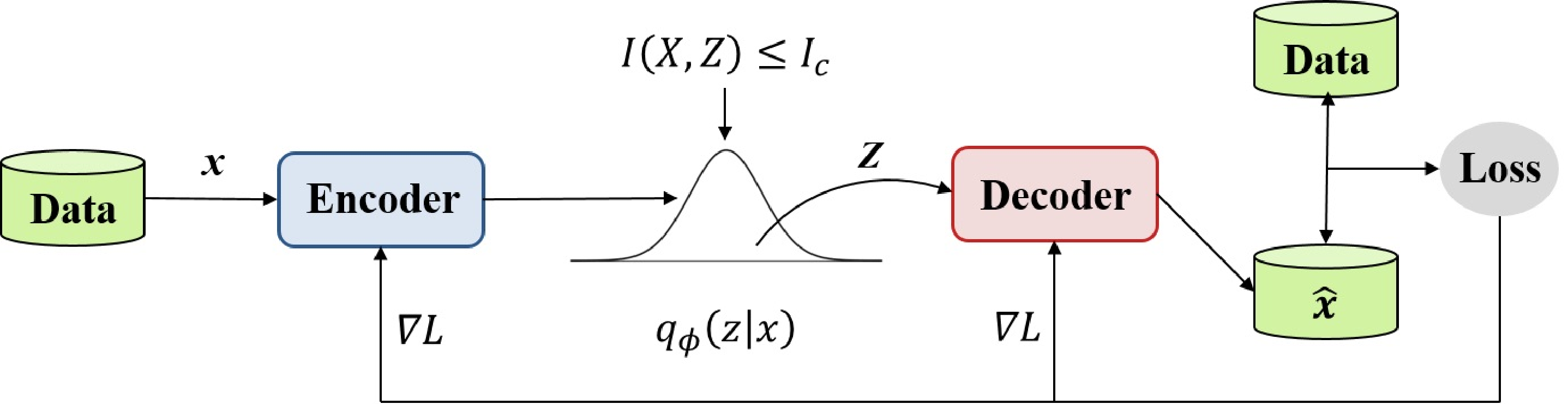}
  \caption{An overview of $\beta$-VAE within the encoder-decoder framework. The encoder maps the input sample $x$ into the latent space subject to the mutual information constraint. The resulting latent sample $z$ satisfies the posterior distribution $q_\phi(z|x)$ usually chosen to be a disentangled multivariate Gaussian in a much lower dimension. The decoder is  trained to map the latent space back to the original input space and generate a decoded sample $\hat{x}$ which is typically ``slimier" or close to the input sample $x$.}
  \label{fig:beta-VAE}
\end{figure}

A direct implementation of the information bottleneck framework with generative modelling is $\beta$-VAE. As illustrated in Fig. \ref{fig:beta-VAE}, $\beta$-VAE  contains two neural networks: an encoder network and a decoder network. The encoder receives the input data $x$ (samples) and tries to find the latent representation $z$ satisfying the mutual information constraint, while the decoder receives $z$ and tries to generate synthetic data $\hat{x}$ to mimic the input data $x$ (reconstruction). In practice, because the mutual information $I$ is hard to calculate, the problem is approximated by minimizing the loss function \cite{Alemi17}:
    \begin{equation}
        \begin{split}
 \mathcal{L}=-\int p_D(x)q_\phi(z|x) \ln p_\theta (x|z)  \, \dd x \dd z \\
 + \beta \frac{1}{N}\sum_{i=1}^N D_{\text{KL}}[q_\phi(z|x_i)\| p(z)],
        \end{split}
        \label{Eq:beta-VAE-lower bound}
    \end{equation}
where $p_D(x)$ is the true data distribution determined by the underlying physical model, $q_\phi(z|x)$ is the posterior probability distribution approximated by the encoder neural network, $p_\theta(x|z)$ is the parameterized data distribution implemented by the decoder, $x_i$ denotes the sample from the dataset, $N$ is the size of the dataset, $p(z)$ is the prior distribution of the latent variable often chosen to be Gaussian, and  $D_{\text{KL}}$ is the so-called Kullbach-Leibler (KL) divergence:
\begin{equation}
D_{\text{KL}}[q_\phi(z|x_i)\| p(z)] = \int \dd z q_\phi(z|x_i) \ln q_\phi(z|x_i)/p(z).
\end{equation}
The posterior $q_{\phi}(z|x)$ is also chosen to be Gaussian to enable the usage of reparameterization trick to avoid high gradient variance \cite{Kingma13}. The detailed form of $p_\theta(x|z)$ depends on the type of the data. As will be introduced later, we use the Bernoulli decoder for binary data, multinomial decoder for non-binary discrete data, and Gaussian decoder for real-valued (continuous) data.

For $\beta = 1$, the above formula recovers the so-called evidence lower bound (ELBO) used in VAEs \cite{Kingma13,Rezende14, Hoffman2016}. The first term refers to the log-likelihood of the data and its value reflects the distortion of the reconstruction. The second term gives the distance of the output $q_\phi(z|x_i)$ and the prior distribution $p(z)$, and its value is zero when no information is encoded. New samples can be generated by running the decoder alone after training. Different choices of the hyperparameter $\beta$ control the trade-off between exact reconstruction (first term) and the extraction of information (second term). The algorithms are hence called $\beta$-VAE \cite{betaVAE}, where one may choose a proper value of $\beta$ to obtain the best disentangled representation $z$.

    \section{Application to the Ising model}
We first apply the above $\beta$-VAE framework to the classical Ising model on a square lattice,
    \begin{equation}
        H = -J\sum_{\langle ij\rangle}S_i S_j + h \sum_i S_i,
    \end{equation}
where $S_i=\pm1$ is the Ising spin, $J$ is the exchange coupling, and $h$ is the magnetic field. The sum over $\langle ij \rangle$ denotes that only the nearest neighbor interaction is considered. For $h=0$, the model undergoes a ferromagnetic (FM) or antiferromagnetic (AFM) phase transition at the critical temperature $T_c/|J| \approx 2.269$ depending on the sign of $J$ \cite{Onsager1944}. For simplicity, we set $|J|=1$ as the energy unit and generate the data set of size $100000$ via Monte Carlo simulations on the square lattice of the size $L= 32$. The algorithm is implemented with Tensorflow \cite{TensorFlow16} using the Adam optimizer \cite{Kingma15}, as elaborated in the Appendix. For better visualization, we set the dimension of the latent variable $z$ to $d_z=2$. We will train the network on the whole dataset including paramagnetic (PM) samples and discuss the quality of their generated samples and the latent representations with respect to the value of $\beta$. For convenience, we will set $h=0$ and use binary data $\sigma_i\in\{0,1\}$ to represent the up/down configuration of the Ising spins. The decoder then takes the Bernoulli form: 
\begin{equation}
p_\theta(x|z)= \prod_{k=1}^{L\times L} m_k^{\sigma_k} (1-m_k)^{1-\sigma_k},
  \label{Eq:Exp-Bernoulli}
\end{equation}
where $x=(\sigma_1,\dots,\sigma_{L\times L})$ is the raw data point for a vectorized spin configuration on the square lattice, and $\{m_k\}$ are Bernoulli parameters as well as the output of the decoder. After training, new synthetic data can be generated by sampling the random variable $z$.

    \begin{figure}[t]
      \centering
      \includegraphics[width=0.4\textwidth]{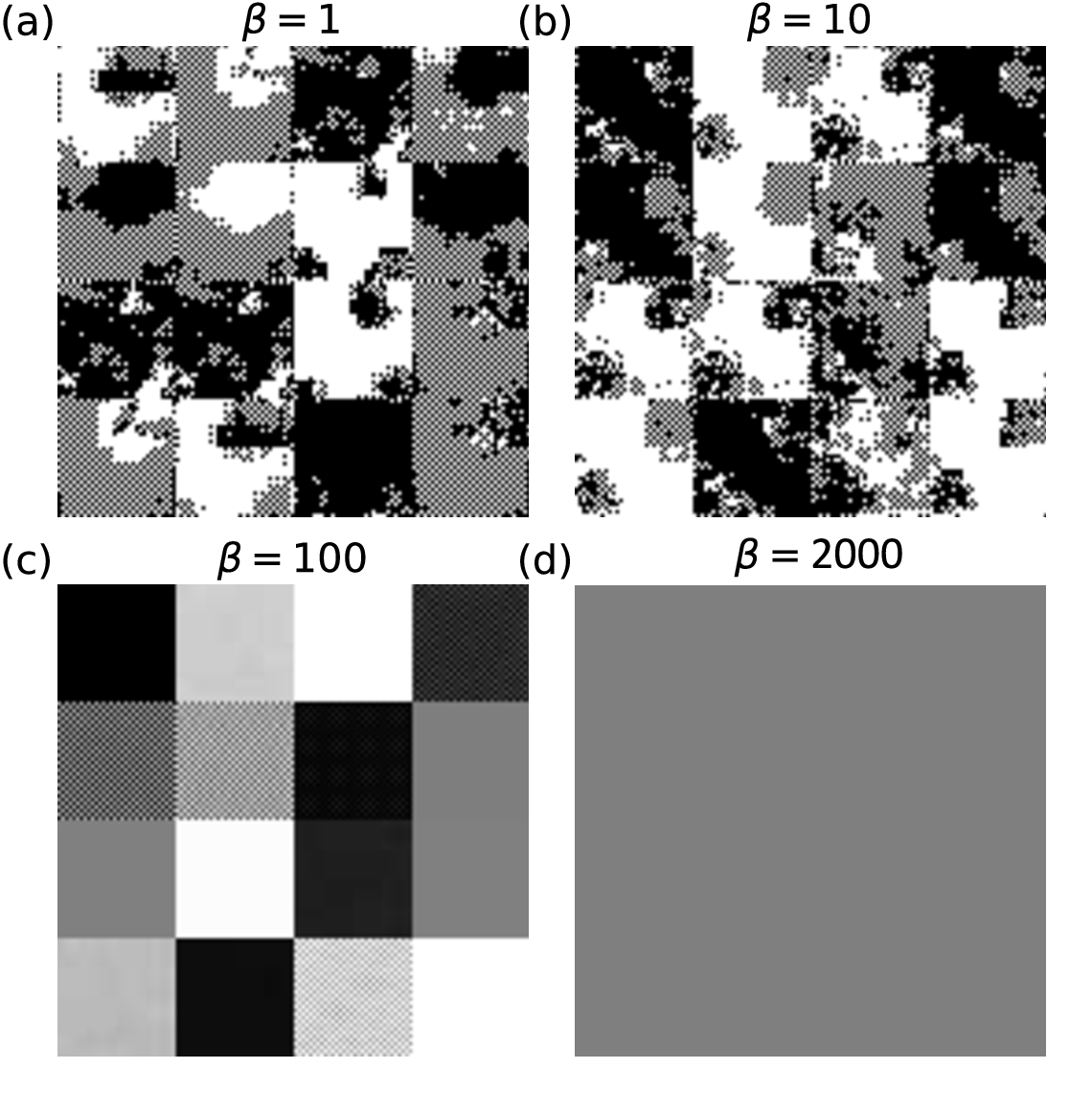}
      \caption{Comparison of typical generated samples for the hyperparameter $\beta=1$, 10, 100, 2000 in three-phase learning. Each small square represents a spin configuration on the $L\times L$ lattice and each panel contains 16 samples. The white, black, gray pixels reflect the learned $m_k$ of roughly 0, 1, 1/2. Hence, the white and black regions correspond to FM phases of opposite polarization, the gray region is PM, and the checkered region is AFM. Only at $\beta=100$, all three phases are well reproduced and separated. At small $\beta$, the PM phase is ignored, while at large $\beta$, the FM and AFM phases are missing.}
      \label{fig:all-generation}
    \end{figure}
    
Figure \ref{fig:all-generation} compares the generated data for different choices of $\beta$. Each small square corresponds to one sampled spin configuration on the $L\times L$ lattice and each panel contains 16 independent samples. Their different patterns correspond to the FM (black and white), AFM (checkered), and PM (gray) phases, respectively. Note that instead of plotting $\{\sigma_k\}$, we take the common practice in the literature and show the Bernoulli parameters $\{m_k\}$. The latter can be viewed as the mean value of the spins in each phase, which allows for a better evaluation of the phase learning. Each $m_k$ should be around $1/2$ in PM, and close to 0 or 1 in AFM (checkerboard) and FM (uniform). Only for $\beta=100$, we see the trained model can correctly generate samples for all three phases, while for usual VAE ($\beta=1$), it fails completely. The reason can be traced back to the loss function given in Eq. \eqref{Eq:beta-VAE-lower bound}, where the two terms try to match the output distribution $p_\theta(x|z)$ with the initial data associated with corresponding $z$ and at the same time match the latent distribution $q_\phi(z|x)$ with the prior distribution $p(z)$, respectively. There is a subtle balance between them, which is controlled by the magnitude of $\beta$. If $\beta$ is small and the $\beta$-term is not properly optimized, $q_\phi(z|x)$ may strongly deviate from the latent distribution and cannot be used to generate correct samples. One the other hand, if $\beta$ is too large so that the first term (likelihood) is bad, the decoder may not be well trained and thus fail to reconstruct the input samples. A proper choice of $\beta$ is therefore required to balance both and produce the best data generation. In this sense, $\beta$ plays the role of regularization to constrain the output range of encoder and decoder correspondingly.

    \begin{figure}[t]
      \centering
      \includegraphics[width=0.45\textwidth]{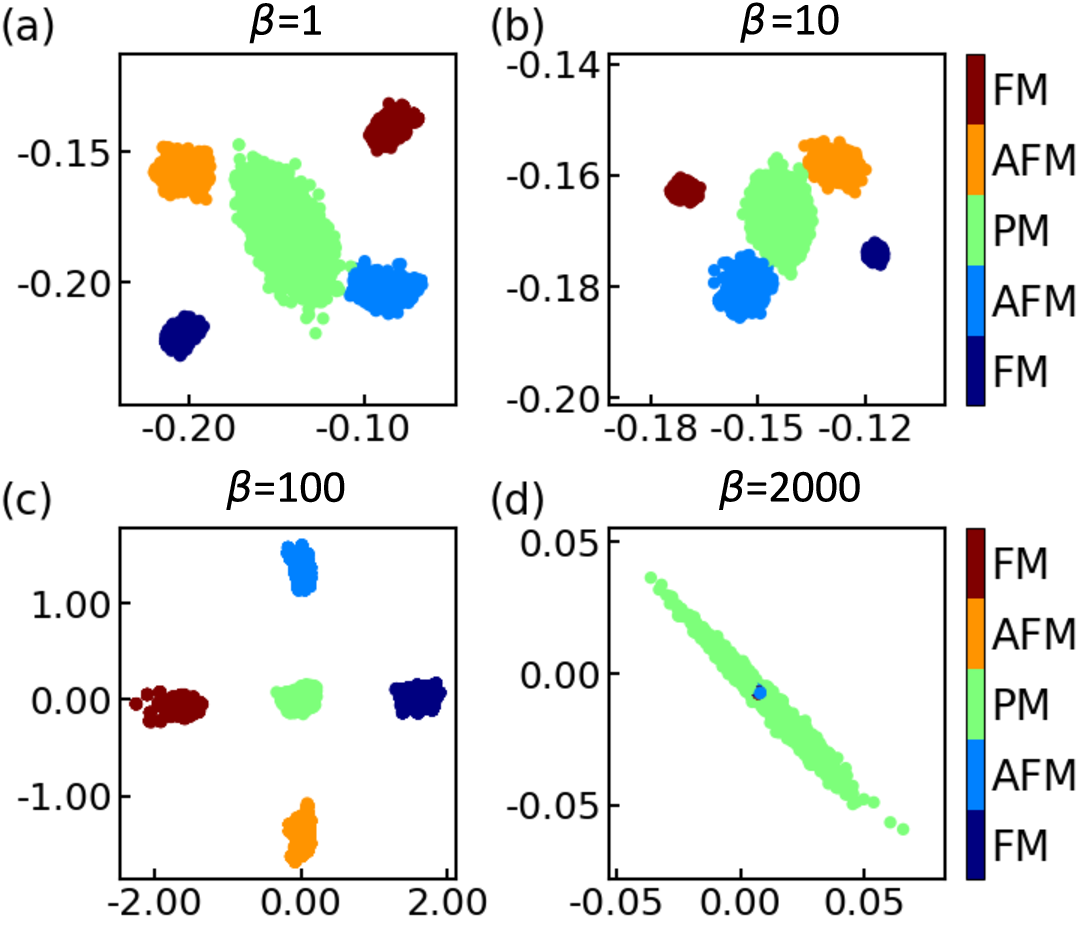}
      \caption{Comparison of the latent representations ($z$) for $\beta=1$, 10, 100, 2000. Each point represents an input sample. Only at $\beta=100$, the disentanglement becomes axis-aligned. For $\beta=2000$, all samples collapse together and become overlapped.}
      \label{fig:representation}
  \end{figure}

Figure \ref{fig:representation} plots the resulting latent representation. We see that the samples are clustered for all three phases already for small $\beta=1$. A larger $\beta$ means that the deviation between $q_\phi(z|x)$ and $p(z)$ is penalized heavier. For ultra large $\beta = 2000$, all samples collapse together in the latent space and $\beta$-VAE fails to distinguish FM and AFM phases from PM. In this case, as shown in Fig. \ref{fig:all-generation}(d), it can only generate PM samples while ignores all features from other phases. Only for $\beta = 100$, the latent representations form a diamond shape with PM samples clustered in the center and FM and AFM samples separated into two portions on two perpendicular axes. In other words, the disentangled representations tend to be axis-aligned for FM and AFM samples. The separation and symmetric clustering on each axis in latent space reflects the $Z_2$ symmetry breaking of both phases. But AFM is also different from FM and contains two sublattices. The $\beta$-VAE correctly captures these properties and creates an independent parameter (the horizontal axis) to distinguish the FM and AFM samples. 

\begin{figure}[t]
  \includegraphics[width=0.45\textwidth]{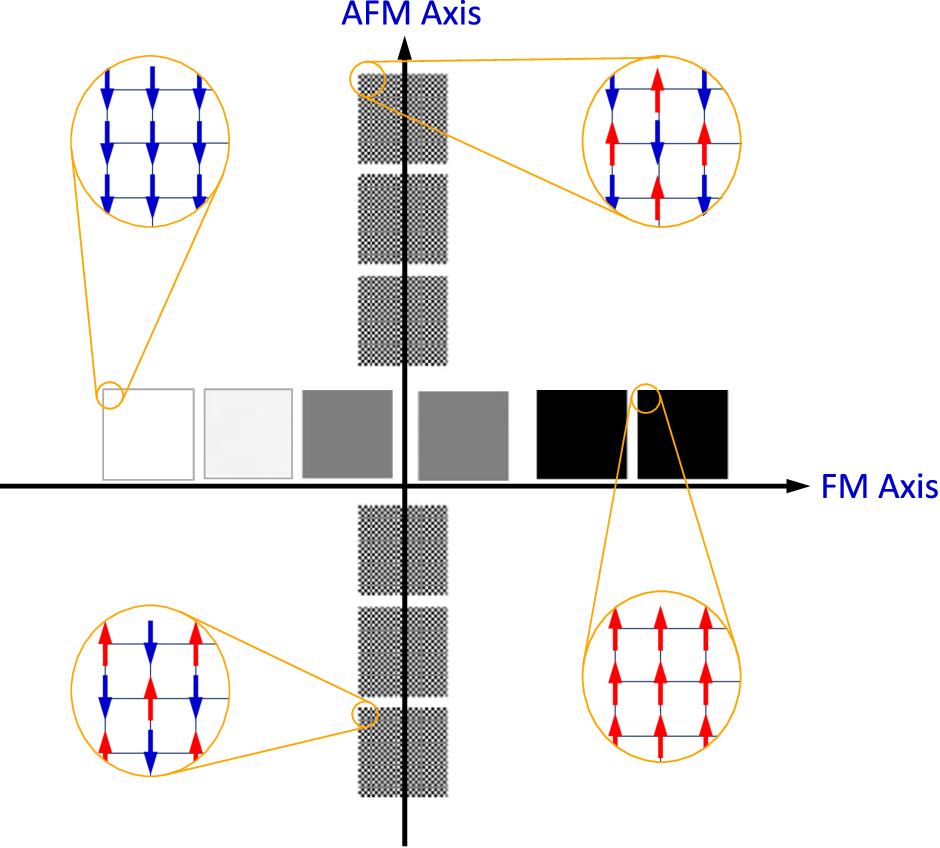}
  \caption{Visualization of deep representations learned by $\beta$-VAE for $\beta=100$. Each axis represents a latent variable that is seen to capture a physical order parameter. The horizontal and vertical axes correspond to the FM and AFM order parameters of the Ising model, respectively.}
  \label{fig:Exploring-Latet-Space}
\end{figure}

The orthogonality of the latent representation $(|z_{\textrm{FM}}^T z_{\textrm{AFM}}|\approx 0)$ indicates different order parameters of the FM and AFM phases. To have a more intuitive visualization of the latent space, we examine the deep learned latent representations in Fig. \ref{fig:Exploring-Latet-Space}. The reconstructed data capture well the main physics of the Ising model. We see a clear consistency between disentangled representation in $\beta$-VAE and the physical order parameters for both FM and AFM phases, where the horizontal axis corresponds to magnetization for FM and the vertical axis gives the staggered magnetization for AFM. $\beta$-VAE generates AFM samples with sublattices depending on the sign of the relevant latent variable. 

The latent space shows certain isometry property of the encoder, namely, the relative distance in input space is protected. To see this, we may view the distribution of FM samples as a mixed distribution, $P = \frac{1}{2}P_- + \frac{1}{2}P_+$, where the subscript denotes the direction of magnetization and the coefficient reflect the $Z_2$ symmetry. Obviously, no similar decomposition exists in the PM phase. We may measure the dissimilarity of different phases via maximum mean discrepancy (MMD) \cite{gretton12a} defined by
\begin{equation}
  \mathrm{MMD}[\mathcal{F},P_x,P_y] \equiv \sup_{f\in \mathcal{F}} \mathrm{E}_{x\sim P_x} [f(x)] - \mathrm{E}_{y\sim P_y} [f(y)],
\end{equation}
where $P_x$ and $P_y$ are the probability distribution of random variables $X$ and $Y$, respectively. A closed form of MMD is usually intractable, so we approximate MMD in the scaling function space $\mathcal{F}=\{f:\mathbb{R}^N \rightarrow \mathbb{R}^N |f(x)=cx,\ 0\leq c \leq 1 \}$. The square of MMD then gives
\begin{align}
     \mathrm{MMD}^2[\mathcal{F},P_x,P_y] &= \sup_{0\leq c \leq 1} c^2(\mathrm{E}_{x\sim P_x} [x] - \mathrm{E}_{y\sim P_y} [y])^2 \nonumber \\
     &= (\mathrm{E}_x [x] - \mathrm{E}_y [y])^2,
\end{align}
which is simply the square of the Euclidean norm of the difference of the first moment. For the input Ising data, the MMD is nothing but the magnetization. For the data in the latent space, it can be calculated straightforwardly from the distribution in Fig.~\ref{fig:representation}(c). We see clearly the equality of distance between $P_+$, $P_-$, and $P_{\text{PM}}$ in both input and latent space, and the isometry property is protected up to a scaling factor. The isometry property of probability plays a crucial role in reflecting the broken symmetry of the Ising model and is not included a priori in the theoretical framework of information bottleneck principle. A careful study of this property may lead to novel understanding of representation learning.

\section{Interpretation of $\beta$-VAE}
How can disentangled representations consist with physical order parameters and exhibit real physical interpretation? We will show below that the Bernoulli decoder corresponds to a mean-field solver and the hyperfine parameter $\beta$ probes the energy difference between learned phases.

\subsection{The Bernoulli decoder as a mean-field solver}
To see this, we first rewrite Eq. \eqref{Eq:Exp-Bernoulli} as
\begin{equation}
  \begin{split}
    p_\theta(x|z) &= \frac{e^{\sum_k h_k \sigma_k}}{\prod_{k}\left(1+e^{h_k}\right)},
  \end{split}
\end{equation}
with $h_k = \ln \frac{m_k}{1-m_k}$ or $m_k = \frac{1}{1+e^{-h_k}}$. This formula corresponds exactly to a mean-field Hamiltonian of the data and its partition function:
\begin{equation}
  H = -\sum_{k} h_k \sigma_k,~~~Z = \text{Tr}e^{-H}= \prod_{k}\left(1+e^{h_k}\right),
  \label{Eq:Mean-Field-Hamiltonian}
\end{equation}
where the parameters $\{h_k\}$ play the role of local effective fields on the lattice and are determined by the decoder neural network and latent variable $z$. Furthermore, the functional relation between $h_k$ and $m_k$ is seen to exactly correspond to the sigmoid activation function $\sigma: \mathbb{R}\rightarrow \mathbb{R}$, $x\rightarrow 1/(1+e^{-x})$ in the last layer, which means that the decoder neural network is learning the local effective field $\{h_k\}$ in the last layer and then uses the activation to give the Bernoulli parameters $\{m_k\}$ as the output. Therefore, activation is equivalent to solving the Hamiltonian \eqref{Eq:Mean-Field-Hamiltonian} with $\{m_k\}$ being the mean-field order parameter of the learned phase, albeit at a fixed temperature $T=1$.

\subsection{The role of the hyperparameter $\beta$}
Given the above correspondence, the role of $\beta$ may then be understood if we rewrite the optimization of the loss function in a stepwise manner. For fixed $p_\theta$, the optimal condition $\delta \mathcal{L}/\delta q_\phi|_{p_\theta}=0$ yields the relation
\begin{equation}
  q_\phi(z|x)=\frac{p(z)p_\theta(x|z)^{1/\beta}}{Z_{\theta,x}},
  \label{Eq:Optimal Encoder}
\end{equation}
where $Z_{\theta,x}=\int dz p(z)p_\theta(x|z)^{1/\beta}$ is the normalization factor. Putting this equation back into the loss function, we find 
$\mathcal{L}=-\beta\langle \ln Z_{\theta,x}\rangle_x$, which is a function of $p_\theta$ solely. Thus, minimizing the loss function in the neural network may also be regarded as minimizing the free energy of a statistical system if we view $Z$ as a partition function of the latent variables. In practice, we find that the encoder $q_\phi$ always converges faster than the decoder $p_\theta$, so that the encoder may be considered as the follower and the decoder as the leader. In this sense, $\beta$-VAE might be viewed as a Stackelberg game \cite{GameTheory}.

To clarify the role of $\beta$, we note that the most widely-used  Bernoulli, multinomial, and Gaussian distributions can all be written in a Boltzmann form with their particular energy functions. As we have seen, the Bernoulli distribution can be derived from an effective Ising Hamiltonian at a fixed temperature. Hence, we may quite generally assume a Boltzmann form for the decoder distribution, $p_\theta(x|z) \propto e^{-E_\theta(x,z)}$. Following Eq.~\eqref{Eq:Optimal Encoder}, the optimal encoder distribution then takes the form, $q_\phi(z|x) \propto e^{-E_\theta(x,z)/\beta}$. This immediately suggests a connection between statistical physics and $\beta$-VAE, with $\beta$ playing the role of an effective temperature in the latent space. This ``temperature" is not a physical temperature, but reflects the overall distribution of the dataset containing different phases. For a well-trained neural network, a proper choice of the hyperparameter $\beta$ must be able to distinguish the energy difference between these phases but in the meanwhile ignore fluctuations within each phase, so that similar samples may cluster in the latent space as demonstrated in Fig. \ref{fig:representation}. If $\beta$ is too large, the $\beta$-VAE cannot distinguish different phases, but if $\beta$ is too small, the neural network may be too sensitive to the fluctuations within each phase. This is consistent with the role of $\beta$ in the loss function (Eq. \eqref{Eq:beta-VAE-lower bound}), where a very large $\beta$ ignores the data information (first term) and a very small $\beta$ pays too much attention to the data details and ignores their global phase feature (second term). One may then give a crude estimate of $\beta$ based on the energy difference of different phases. For example, in the Ising model considered here, the energy difference between FM (AFM) and PM is given by $|J|L^2/2$, which is about $512$ in our calculations for $|J|=1$ and $L=32$. Hence, a proper choice of $\beta$ should be about a fraction of $512$. Indeed, as shown in Fig. \ref{fig:representation}, a larger $\beta=2000$ fails to learn anything, while a smaller $\beta=10$ cannot capture correct phases. Only for a proper value of $\beta=100$, the $\beta$-VAE network correctly identifies all three phases and generates good samples.

\section{Generalization to other models} 
Our interpretation of the $\beta$-VAE is not limited to the Ising model. To show its applicability in general cases, we extend it to more complex classical models with non-binary discrete data or real-valued continuous data. Depending on the type of the data, we introduce different decoder neural networks.

\subsection{Multinomial decoder for non-binary discrete data}
\begin{figure}[t]
  \centering
  \includegraphics[width=0.45\textwidth]{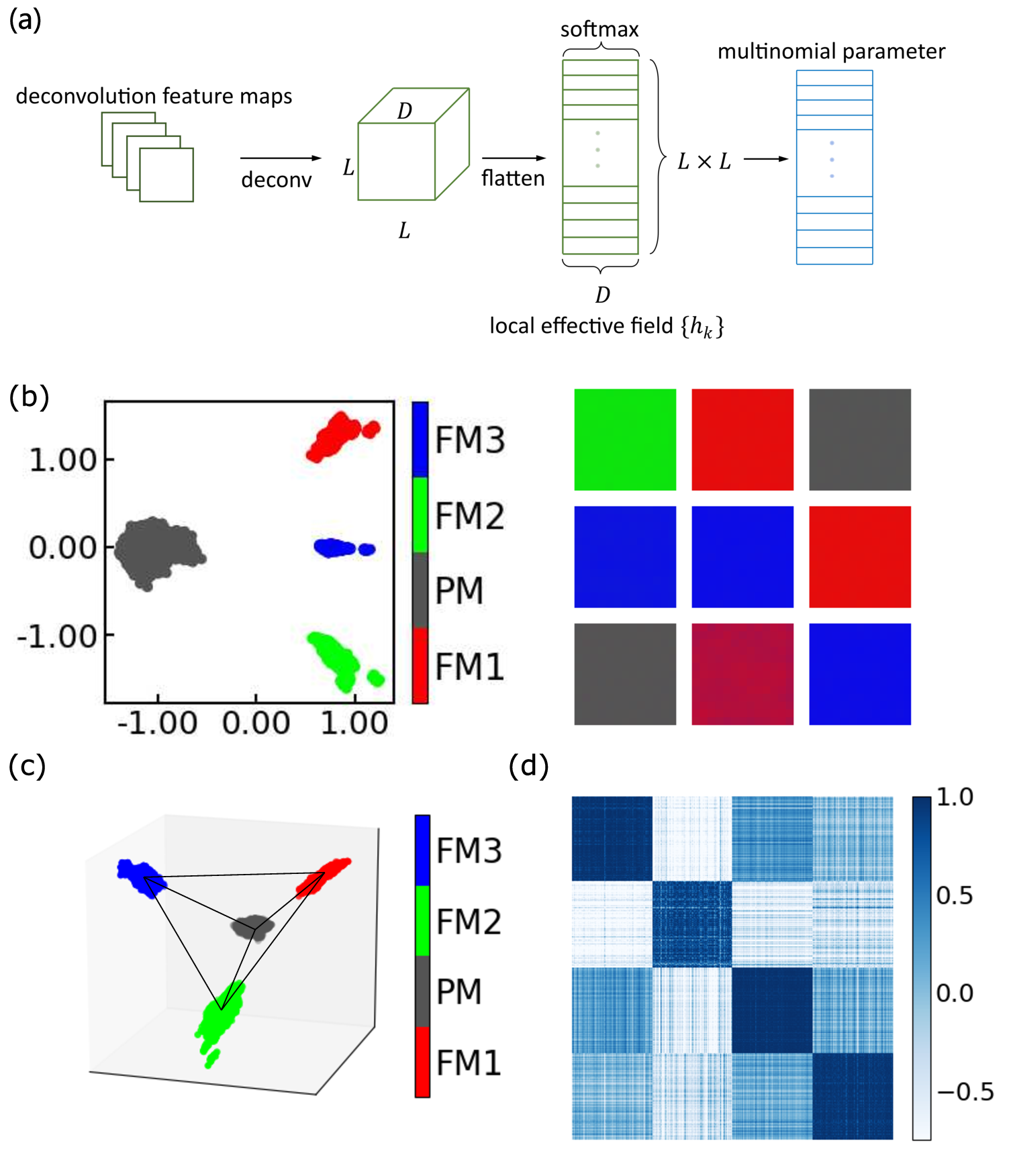}
  \caption{(a) Architecture of the multinomial decoder. The output of the decoder has $D$ channels and a channel-wise softmax function is used for activation. (b) The latent representation and generated samples for $d_z=2$ and $\beta = 65$. (c) The latent representation for $d_z = 3$ and $\beta = 70$. (d) The cosine similarity of the latent representation for $d_z = 4$ and $\beta = 71$, where the axes represent the data points in the latent space and each pixel represents the normalized inner product of two data points. A synthetic dataset was used for the test as described in the main text.}
  \label{Fig:multinomial decoder}
\end{figure}

We consider here a system of classical spins taking $D$ discrete values. These spin values can be mapped to a $D$-dimensional vector with only one non-zero entry. For $D=3$, the spin $\sigma^k_d$ takes values in $\{-1,0,1\}$ and can be represented as $(1,0,0)$, $(0,1,0)$, $(0,0,1)$, respectively. A multinomial decoder can then be used with the form:
\begin{equation}
  p_\theta(x|z) = \prod_{k=1}^{L\times L}\prod_{d=1}^D m_{kd}^{\sigma^k_d},
  \label{Eq:multinomial decoder}
\end{equation}
where the multinomial parameter $\{m_{kd}\}_{d=1}^D$ is the output of the decoder satisfying the constraint $\sum_{d=1}^D m_{kd}= 1$ for $k=1,\dots,L\times L$, and $\sigma^k$ is the classical spin in the form of a $D$-dimensional vector on the $k$-th lattice point. An illustration of its architecture is given in Fig.~\ref{Fig:multinomial decoder}(a). If we use the channel-wise softmax activation function in the last layer of the decoder neural network,
\begin{equation}
  m_{kd} = \frac{e^{h^k_d}}{\sum_{d'} e^{h^k_{d'}}},
  \label{Eq:softmax}
\end{equation} 
$p_\theta$ can be rewritten again in the Boltzmann form:
\begin{equation}
  p_\theta(x|z) = \frac{e^{\sum_{k,d} h^k_d \sigma^k_d}}{\prod_{k} \left(\sum_{d} e^{h^k_d}\right)},
\end{equation}
so that we reach a mean-field model:
\begin{equation}
  H = -\sum_{k=1}^{L\times L} \sum_{d=1}^D h^k_d \sigma^k_d, ~~~Z = \prod_{k=1}^{L\times L}\left(\sum_{d=1}^D e^{h^k_d}\right),
  \label{Eq: Mean-field non-binary}
\end{equation}
where $\{h^k_d\}$ is nothing but the ``local effective field", and the activation function \eqref{Eq:softmax} gives the mean-field solution $\{m_{kd}\}$. Compared to the Bernoulli decoder in the Ising case, both the local effective fields and the spins now take a vector form rather than a scaler.

To show the validity of this neural network, we apply it to a synthetic dataset of $D=3$ generated using the multinomial parameters $m_{kd}^0=(1/3,1/3,1/3)$, $(0.9, 0.05, 0.05), (0.05, 0.9, 0.05)$, $(0.05, 0.05, 0.9)$, which mimick the disordered phase and three polarized phases that break the $Z_3$ symmetry, respectively. After training, the latent representation of the dataset and the generated samples (the output $m_{kd}$) are shown in Fig. \ref{Fig:multinomial decoder}(b) for $\beta=65$ with the latent dimension $d_z=2$. We see clear separation and clustering of all four phases in the latent space, with samples of the disordered phase (PM) located on a horizontal axis, and those of three polarized phases (FM) distributed on a vertical line. The generated samples are also of good quality and well captures the feature of each phase.

To explore the role of the latent dimension, we also plot the latent representation for $d_z=3$ and $\beta=70$ in Fig. \ref{Fig:multinomial decoder}(c). The data are seen to cluster on four vertices of a tetrahedron. Interestingly, we find a similar distance of all three polarized phases from the PM, which retains the $Z_3$ symmetry of the underlying model. A straightforward comparison with the latent distribution for $d_z=2$ suggests that a better representation may be achieved by tuning the dimension of the latent space. However, for  $d_z>3$, a direct visualization of the latent distribution is impossible, so we introduce the cosine similarity defined as the inner product of two normalized vectors of the data to characterize the distribution in the latent space. A similarity matrix is plotted in Fig. \ref{Fig:multinomial decoder}(d) for $d_z=4$ and $\beta=71$. The approximate block diagonalization of the matrix indicates the approximate disentanglement of different phases. In all cases, our $\beta$-VAE with multinomial decoder can well distinguish all four phases in the dataset.

\subsection{Gaussian decoder for real-valued continuous data}

\begin{figure}[b]
  \centering
  \includegraphics[width=0.48\textwidth]{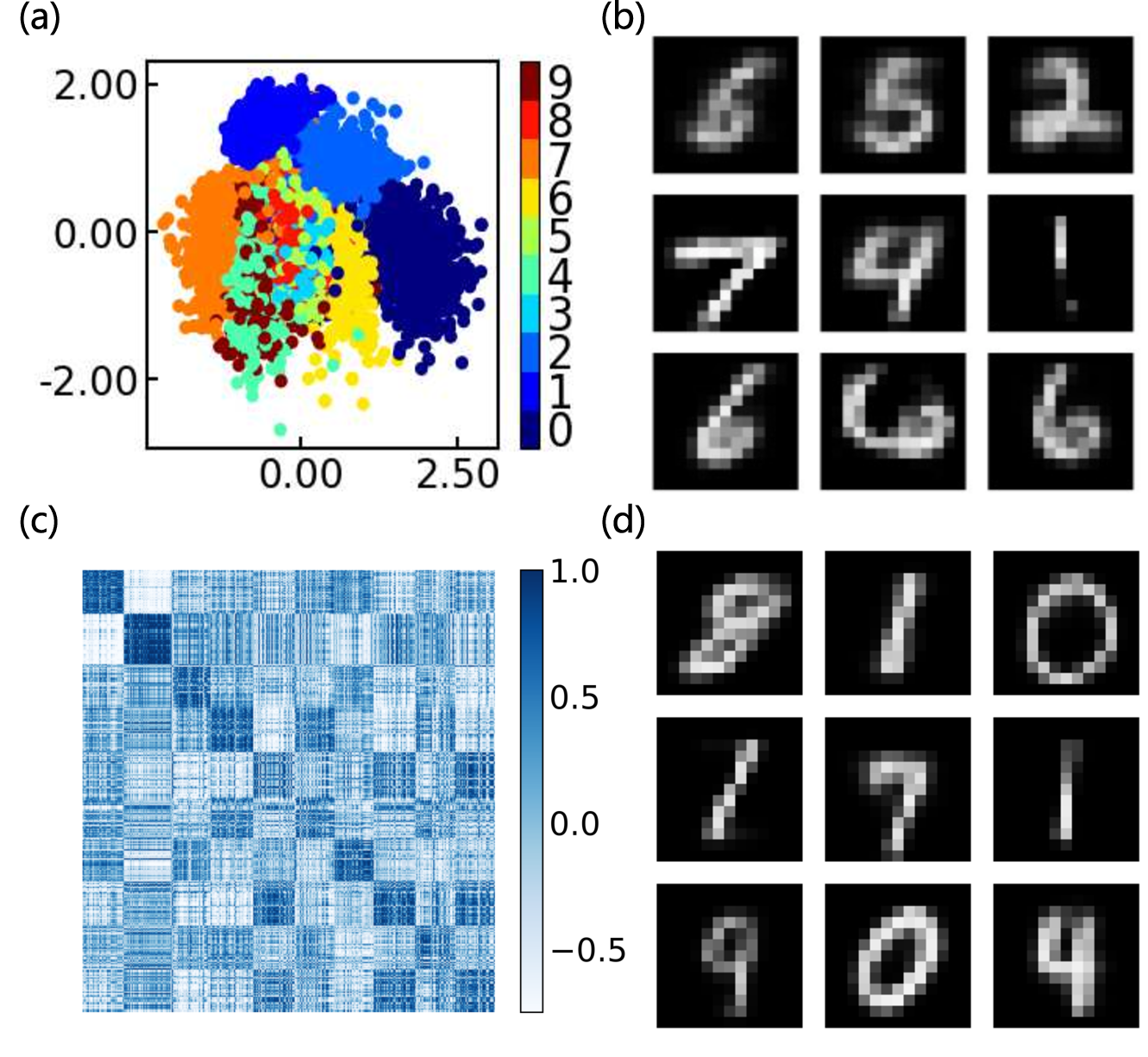}
  \caption{The latent representation and generated samples using $\beta$-VAE with the Gaussian decoder on the MNIST dataset for (a,b) $d_z=2$ and (c,d) $d_z = 16$.}
  \label{fig:MNIST-Gaussian}
\end{figure}

The $\beta$-VAE network can also be applied to classical models with real-valued continuous data $x = (x_1,\dots,x_{L\times L})$ by using the standard Gaussian decoder:
\begin{equation}
  p_\theta(x|z) = \prod_{k=1}^{L\times L}\frac{1}{\sqrt{2\pi}} e^{-(x_k-\mu_k)^2/2}.
  \label{Eq:Gaussian Decoder}
\end{equation}
where $x_k$ is the $k$-th entry and $\mu_k$ is the mean of the Gaussian distribution depending on the hidden variable $z$. The neural network directly yields the output $\mu_k$ and no activation function is needed in the last layer. Similarly, the Gaussian decoder can be viewed as solving a mean-field Hamiltonian of the form:
\begin{equation}
  H = \sum_{k}H_k, ~~~H_k = -\mu_k x_k + \frac{1}{2}x_k^2,
  \label{Eq:Mean Field Gaussian}
\end{equation}
where we have dropped the constant $\mu_k^2/2$.

We tested the Gaussian decoder on the Modified National Institute of Standards and Technology (MNIST) dataset of handwritten digits, where the ordered phases correspond to ten handwritten numbers 0, 1, 2, $\cdots$, 9. The latent representation for $d_z=2$ is plotted in Fig. \ref{fig:MNIST-Gaussian}(a), where the images corresponding to the same number are seen to cluster together but hard to separate due to the low dimensionality of the latent space. For comparison, we further calculated the cosine similarity for a larger $d_z=16$. The resulting matrix is plotted in Fig. \ref{fig:MNIST-Gaussian}(c) and divided into clear blocks. The dark blue diagonal blocks reflect the clustering of different handwritings for the same numbers, while the blue off-diagonal blocks indicate the similarity of the handwritings for two different numbers, such as numbers 9 and 4 shown in Fig. \ref{fig:MNIST-Gaussian}(d). The white off-diagonal blocks on the top left of the matrix suggest that the handwritings of 0 and 1 are well distinguishable. Interestingly, as plotted in Figs. \ref{fig:MNIST-Gaussian}(b) and \ref{fig:MNIST-Gaussian}(d), the generated samples by the decoder are meaningful for both $d_z$.

\section{Discussion and conclusions}
We have studied disentangled low-dimensional representations for the Ising model using the information bottleneck method. Compared to traditional variational autoencoder, $\beta$-VAE can generate samples of appropriate physical properties with proper hyperparameter setting. The latent representation can become axis-aligned and reveal certain isometry property of the encoder. Our results confirm the consistency between learned disentangled low-dimensional representations and FM/AFM order parameters for the Ising model. We then show the validity of our method by applying it to more general models with non-binary discrete or real-valued continuous data using multinomial or Gaussian decoders. These already cover a wide spectrum of physical models including both classical models and quantum models that can be simulated using Monte Carlo by introducing classical auxiliary variables. For example, in the determinant quantum Monte Carlo (DQMC) method, the Hubbard or periodic Anderson models are simulated in the auxiliary Ising configuration space after the Hubbard-Stratonovich transformation \cite{Assaad2008,Wei2017,Hu2019}. Application of our method to quantum models will be explored in future work.

\begin{figure}[t]
  \includegraphics[width = 0.48\textwidth]{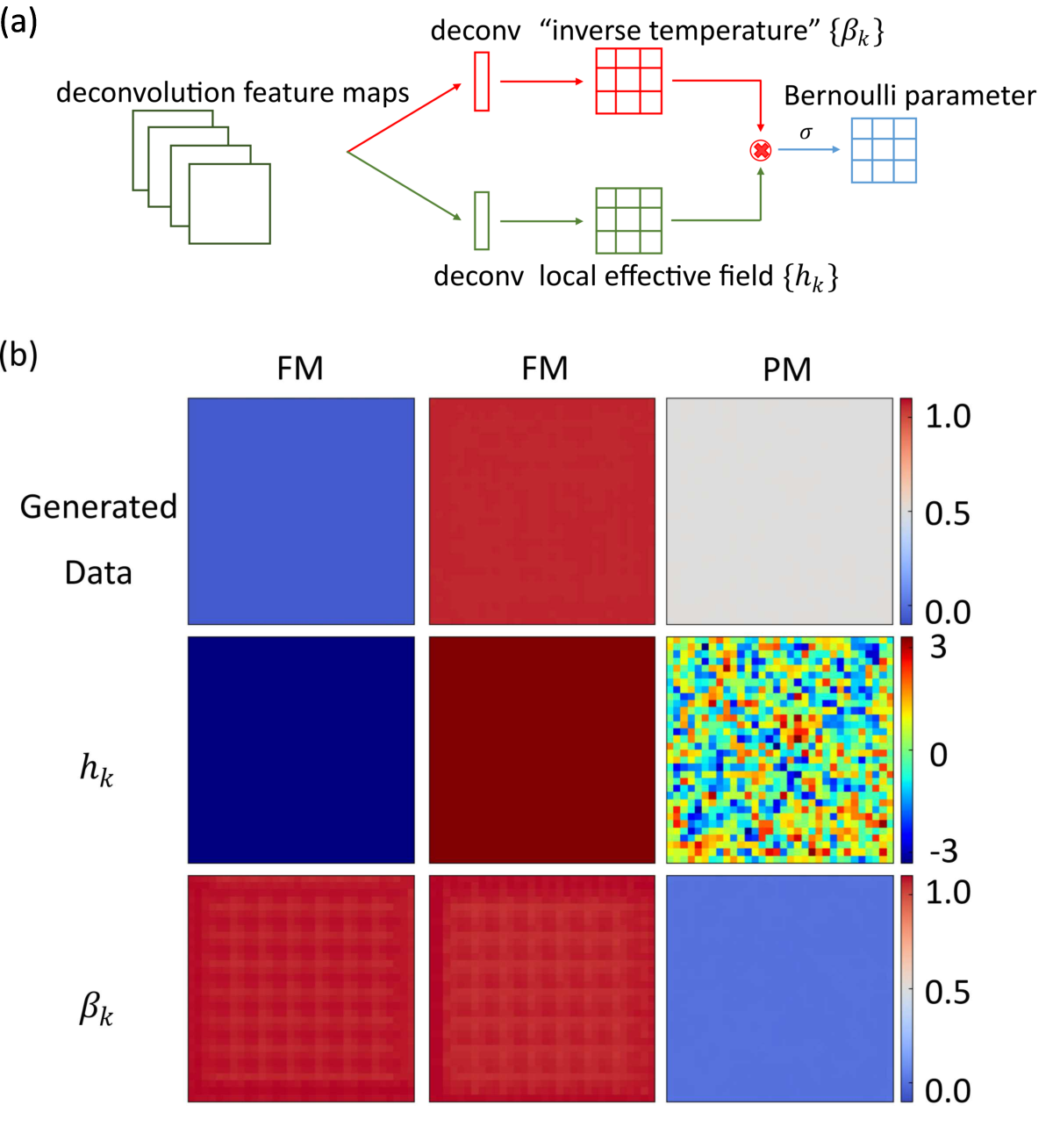}
  \caption{(a) Illustration of the $\beta^2$-VAE architecture, with both the ``temperature branch" $\beta_k$ and the original branch learning a local effective field $h_k$. The symbol $\otimes$ denotes element-wise matrix multiplication. The added branch enables the model to generate samples with thermal fluctuations. (b) Comparison of generated samples for FM and PM phases and their local effective field $h_k$ and ``inverse temperature" $\beta_k$ learned by $\beta^2$-VAE for $\beta=100$ with the constraint $0<\beta_k<1$.}
  \label{Fig:beta2-VAE}
\end{figure}

Another important observation is that all three widely-used decoders can be viewed as solving a mean-field Hamiltonian at a fixed temperature $T=1$. This indicates that the traditional decoder structure lacks a temperature scale and is therefore unable to represent thermal fluctuations in data generation. We may thus consider to improve the decoder by introducing an extra reweighing branch to detect fluctuations. For the Bernoulli decoder, this can be done by replacing the usual sigmoid activation function by $\sigma:h_k,\beta_k \rightarrow 1/(1+e^{-\beta_k h_k})$. As illustrated in Fig. \ref{Fig:beta2-VAE}(a), the new neural network contains two branches: one corresponds to ``local temperature" $\{\beta_k\}$ (red), and the other reflects the effective local field $\{h_k\}$ (green). Both parameters are learned jointly during the training procedure. The sigmoid function is activated only after the element-wise product of $\{\beta_k\}$ and $\{h_k\}$. The temperature branch can also be viewed as a special case of attention mechanism \cite{Vaswani17, parikh16, cheng16}. We name this physics-informed algorithm as $\beta^2$-VAE since $\beta$ is also a standard usage of statistical physics for the true inverse temperature.

As a proof of concept, we have performed one-phase learning with $\beta^2$-VAE by training it with samples from each phase alone. The implementation details can be found in the appendix. Figure \ref{Fig:beta2-VAE}(b) shows the generated samples, learned local effective fields $h_k$, and local ``inverse temperatures" $\beta_k$. For simplicity, $\beta_k$ is valued in $(0,1)$. We find a larger $\beta_k$ (low temperature) in FM and a smaller $\beta_k$ (high temperature) in PM, in good agreement with physical intuition. The generated data (first row) confirm the power of reproducing correct samples within $\beta^2$-VAE. Given a generated sample, the corresponding local effective fields $h_k$ (second row) successfully discover the spontaneous magnetization in the FM phase and disorder in the PM phase. Further implementation of this idea requires more elaborated data analyses.

\acknowledgements
This work was supported by the National Natural Science Foundation of China (Grant No. 11974397) and the Strategic Priority Research Program of the Chinese Academy of Sciences (Grant No. XDB33010100).\\

\appendix
\section{Implementation Details}
Standard CNN architecture has been used for both $\beta$-VAE and $\beta^2$-VAE as listed in Tables \ref{Tab:beta} and \ref{Tab:beta2}, respectively. Both the encoder and decoder contain four layers, but $\beta^2$-VAE has two branches in the last layer. All $\alpha$ in leaky-ReLU (lReLU) are set to $0.2$ and the batch size is set to $128$. Here ReLU denotes linear rectification function and BN stands for batch normalization \cite{ioffe15}. We have used the Adam optimizer with learning rate equal to 0.001, $\beta_1=0.5$, and $\beta_2=0.999$.

    \begin{table}[h]
      \caption{Neural network architecture for $\beta$-VAE. ReLU denotes linear rectification function and BN stands for batch normalization.}
    
    \begin{tabular}{l}
      \hline\noalign{\smallskip}
  
      Encoder network \\
      
      \noalign{\smallskip}\hline\noalign{\smallskip}
      Input gray image $x\in \mathbb{R}^{32 \times 32 \times 1}$ \\
      Conv2d, BN, $4 \times 4 \times 64$, stride=2, padding=SAME, lReLU \\
      Conv2d, BN, $4 \times 4 \times 128$, stride=2, padding=SAME, lReLU \\
      FC, BN, $1024$ lReLU \\
      FC, $2$, linear; FC, $2$, softplus\\       
      \noalign{\smallskip}\hline
      Decoder network \\
      \noalign{\smallskip}\hline\noalign{\smallskip}
      Input $z\in \mathbb{R}^2$ \\
      FC, BN, 1024, ReLU\\
      FC, BN, $8 \times 1024$, ReLU\\
      Deconv2d, BN, $4 \times 4 \times 64$, stride=2, padding=SAME, ReLU\\
      Deconv2d, $4 \times 4 \times 1$, stride=2, padding=SAME, sigmoid\\
      \noalign{\smallskip}\hline
    \end{tabular}
    \label{Tab:beta}
  \end{table}

  \begin{table}[h]
    \caption{Neural network architecture for $\beta^2$-VAE.}
  
  \begin{tabular}{l}
    \hline\noalign{\smallskip}

    Encoder network \\
    
    \noalign{\smallskip}\hline\noalign{\smallskip}
    Input gray image $x\in \mathbb{R}^{32 \times 32 \times 1}$ \\
    Conv2d, BN, $4 \times 4 \times 64$, stride=2, padding=SAME, lReLU \\
    Conv2d, BN, $4 \times 4 \times 128$, stride=2, padding=SAME, lReLU \\
    FC, BN, $1024$ lReLU\\
    FC, $2$, linear; FC, $2$, softplus  \\
    \noalign{\smallskip}\hline
    Decoder network \\
    \noalign{\smallskip}\hline\noalign{\smallskip}
    Input $z\in \mathbb{R}^2$ \\
    FC, BN, 1024, ReLU\\
    FC, BN, $8 \times 1024$, ReLU\\
    Deconv2d, BN, $4 \times 4 \times 64$, stride=2, padding=SAME, ReLU\\
    ~~~\begin{tabular}{l}
      Deconv2d, $4 \times 4 \times 1$, stride=2, padding=SAME, sigmoid\footnote{Blank space indicates two branches in the same layer.} \\ 
      Deconv2d, $4 \times 4 \times 1$, stride=2, padding=SAME, linear
      \end{tabular}\\
      sigmoid \\
    \noalign{\smallskip}\hline

  \end{tabular}
  \label{Tab:beta2}
\end{table}


\begin{thebibliography}{10}

  \bibitem{7780459}
  K.~{He}, X.~{Zhang}, S.~{Ren}, and J.~{Sun}, Deep residual learning for image
    recognition, in {\em 2016 IEEE Conference on Computer Vision and Pattern
    Recognition (CVPR)}, (IEEE, 2016).
  
  \bibitem{NIPS2012_4824}
  A.~Krizhevsky, I.~Sutskever, and G.~E. Hinton, Imagenet classification with
    deep convolutional neural networks, in {\em Advances in Neural Information
    Processing Systems}, (Curran Associates, Inc., 2012).
  
  \bibitem{10.1007/978-3-319-46493-0_38}
  K.~He, X.~Zhang, S.~Ren, and J.~Sun, Identity mappings in deep residual
    networks, in {\em European Conference on Computer Vision}, (Springer International Publishing, 2016).
  
  \bibitem{mnih2013playing}
  V.~Mnih, K.~Kavukcuoglu, D.~Silver, A.~Graves, I.~Antonoglou, D.~Wierstra, and
    M.~Riedmiller, Playing atari with deep reinforcement learning, in {\em
    NIPS Deep Learning Workshop}, (2013).
  
  \bibitem{Schrittwieser2019MasteringAG}
  J.~Schrittwieser, I.~Antonoglou, T.~Hubert, K.~Simonyan, L.~Sifre, S.~Schmitt,
    A.~Guez, E.~Lockhart, D.~Hassabis, T.~Graepel, T.~P. Lillicrap, and
    D.~Silver, Mastering atari, go, chess and shogi by planning with a learned
    model, {\em ArXiv preprint arXiv:1911.08265}, (2019).
  
  \bibitem{AlphaStar}
  O.~Vinyals, I.~Babuschkin, W.~M. Czarnecki, M.~Mathieu, P.~Georgiev, J.~Oh,
    D.~Horgan, M.~Kroiss, I.~Danihelka, A.~Huang, L.~Sifre, T.~Cai, J.~P.
    Agapiou, M.~Jaderberg, A.~S. Vezhnevets, R.~Leblond, T.~Pohlen, V.~Dalibard,
    D.~Budden, Y.~Sulsky, J.~Molloy, T.~L. Paine, C.~Gulcehre, Z.~Wang, T.~Pfaff,
    Y.~Wu, R.~Ring, D.~Yogatama, D.~W{\"{u}}nsch, K.~Mckinney, O.~Smith,
    T.~Schaul, T.~Lillicrap, K.~Kavukcuoglu, D.~Hassabis, C.~Apps, and D.~Silver,
    ``Grandmaster level in Starcraft II using multi-agent reinforcement
    learning,  Nature {\bf 575}, 350 (2019).

  \bibitem{Locatello19}
  F.~Locatello, S.~Bauer, M.~Lucic, G.~Raetsch, S.~Gelly, B.~Sch{\"o}lkopf, and
  O.~Bachem, Challenging common assumptions in the unsupervised learning of
  disentangled representations, in {\em Proceedings of the 36th International
  Conference on Machine Learning}, (PMLR, 2019).

  
  \bibitem{6472238}
  Y.~{Bengio}, A.~{Courville}, and P.~{Vincent}, Representation learning: A
    review and new perspectives, {\em IEEE Trans. Pattern Anal. Mach. Intell.} {\bf  35}, 1798 (2013).
  
\bibitem{pmlr-v80-kim18b}
H.~Kim and A.~Mnih, Disentangling by factorising, in {\em Proceedings of
  the 35th International Conference on Machine Learning}, (PMLR, 2018).
  
  \bibitem{DLYGG}
  Y.~Bengio, Y.~Lecun, and G.~Hinton, Deep learning,  Nature {\bf 521},
    436 (2015).
  
  \bibitem{betaVAE}
  C.~P. Burgess, I.~Higgins, A.~Pal, L.~Matthey, N.~Watters, G.~Desjardins, and
    A.~Lerchner, Understanding disentangling in beta-vae, in {\em Workshop on
    Learning Disentangled Representations at the 31st Conference on Neural
    Information Processing Systems}, (2017).

  \bibitem{goodfellow2014generative}
  I.~Goodfellow, J.~Pouget-Abadie, M.~Mirza, B.~Xu, D.~Warde-Farley, S.~Ozair,
    A.~Courville, and Y.~Bengio, Generative adversarial nets, in {\em
    Advances in Neural Information Processing Systems}, (Curran Associates, Inc., 2014).
  
  \bibitem{chen16}
  X.~Chen, Y.~Duan, R.~Houthooft, J.~Schulman, I.~Sutskever, and P.~Abbeel,
    ``InfoGAN: Interpretable representation learning by information maximizing
    generative adversarial nets, in {\em Advances in Neural Information
    Processing Systems}, (Curran Associates, Inc., 2016).
  
  \bibitem{Kingma13}
  D.~P. Kingma and M.~Welling, Autoencoding variational bayes, in {\em
    Proceedings of the International Conference on Learning Representations},
    (Openreview, 2014).
  
  \bibitem{Rezende14}
  D.~J. {Rezende}, S.~{Mohamed}, and D.~{Wierstra}, Stochastic backpropagation
    and approximate inference in deep generative models, in {\em Proceedings of
    the 31th International Conference on Machine Learning}, (PMLR, 2014).
  
  \bibitem{Smolensky1986}
  P.~Smolensky, {\em Information processing in dynamical systems: Foundations of harmony theory.} (MITPress, 1986).

  \bibitem{Amin2016QuantumBM}
  M.~H. Amin, E.~Andriyash, J.~Rolfe, B.~Kulchytskyy, and R.~Melko, Quantum
    Boltzmann machine, Phys. Rev. X {\bf 8}, 021050 (2018).

  \bibitem{Koch-Janusz2018}
  M.~Koch-Janusz and Z.~Ringel, {Mutual information, neural networks and the
    renormalization group}, Nat. Phys. {\bf 14}, 578 (2018).

  \bibitem{Nagy2019}
  A.~Nagy and V.~Savona, Variational quantum Monte Carlo method with a
    neural-network ansatz for open quantum systems, Phys. Rev. Lett.
    {\bf 122}, 250501 (2019).

  \bibitem{Hartmann2019}
  M.~J. Hartmann and G.~Carleo, Neural-network approach to dissipative quantum
    many-body dynamics, Phys. Rev. Lett. {\bf 122}, 250502 (2019).

  \bibitem{Vicentini2019}
  F.~Vicentini, A.~Biella, N.~Regnault, and C.~Ciuti, Variational
    neural-network ansatz for steady states in open quantum systems, Phys. Rev. Lett. {\bf 122}, 250503 (2019).

  \bibitem{Yoshioka2019}
  N.~Yoshioka and R.~Hamazaki, Constructing neural stationary states for open
    quantum many-body systems, Phys. Rev. B {\bf  99}, 214306 (2019).

  \bibitem{Torlai2016}
  G.~Torlai and R.~G. Melko, Learning thermodynamics with Boltzmann machines, 
      Phys. Rev. B {\bf  94}, 165134 (2016).

  \bibitem{Torlai2018}
  G.~Torlai, G.~Mazzola, J.~Carrasquilla, M.~Troyer, R.~Melko, and G.~Carleo,
    ``{Neural-network quantum state tomography}, Nat. Phys. {\bf 14},
    447 (2018).

  \bibitem{Torlai2018b}
  G.~Torlai and R.~G. Melko, Latent space purification via neural density
    operators, Phys. Rev. Lett. {\bf 120}, 240503 (2018).


  \bibitem{Carleo2017}
  G.~Carleo and M.~Troyer, Solving the quantum many-body problem with
    artificial neural networks,  Science {\bf 355}, 602 (2017).
  
  \bibitem{Alan2018}
  A.~Morningstar and R.~G. Melko, Deep learning the Ising model near
    criticality, J. Mach. Learn. Res. {\bf  18},
    5975 (2018).
  
  \bibitem{DAngelo2020LearningTI}
  F.~D'Angelo and L.~B{\"o}ttcher, Learning the Ising model with generative
    neural networks, {\em ArXiv preprint: arXiv:2001.05361}, (2020).
  
  \bibitem{Efthymiou2019}
  S.~Efthymiou, M.~J.~S. Beach, and R.~G. Melko, Super-resolving the Ising
    model with convolutional neural networks,  Phys. Rev. B {\bf  99},
    075113 (2019).
  
  \bibitem{scholkopf1997}
  B.~Sch{\"o}lkopf, A.~Smola, and K.-R. M{\"u}ller, Kernel principal component
    analysis, in {\em Artificial Neural Networks --- ICANN'97}, (Springer, 1997).
  
  \bibitem{Wang2016}
  L.~Wang, Discovering phase transitions with unsupervised learning, Phys. Rev. B, {\bf  94}, 195105 (2016).
  
  \bibitem{Li2018}
  S.-H. Li and L.~Wang, Neural network renormalization group, Phys. Rev. Lett. {\bf 121}, 260601 (2018).
    
  \bibitem{Iso2018}
  S.~Iso, S.~Shiba, and S.~Yokoo, Scale-invariant feature extraction of neural
    network and renormalization group flow, Phys. Rev. E {\bf  97},
    053304 (2018).
  
  \bibitem{Kim2018}
  D.~Kim and D.-H. Kim, Smallest neural network to learn the Ising
    criticality, Phys. Rev. E {\bf  98}, 022138 (2018).
  
  \bibitem{Gao2018AQM}
X.~Gao, Z.~Zhang, and L.-M. Duan, A quantum machine learning algorithm based
  on generative models, Sci. Adv. {\bf 4}, eaat9004 (2018).

  \bibitem{Lloyd2018}
  S.~Lloyd and C.~Weedbrook, Quantum generative adversarial learning, Phys. Rev. Lett. {\bf 121}, 040502 (2018).

  \bibitem{Zoufal2019}
C.~Zoufal, A.~Lucchi, and S.~Woerner, {Quantum Generative Adversarial
  Networks for learning and loading random distributions}, npj Quantum Inf. {\bf 5}, 103 (2019). 
   
  \bibitem{gretton12a}
  A.~Gretton, K.~M. Borgwardt, M.~J. Rasch, B.~Sch{{\"o}}lkopf, and A.~Smola, A
    kernel two-sample test, J. Mach. Learn. Res.
    {\bf  13}, 723 (2012).

  \bibitem{Tishby99}
  N.~Tishby, F.~C.~Pereira, and W.~Bialek, The information bottleneck method, {\em arXiv preprint physics/0004057, (2000).}
  
  \bibitem{Shwartz-Ziv17}
  R.~Shwartz-Ziv and N.~Tishby, Opening the black box of deep neural networks
    via information, {\em ArXiv preprint arXiv:1703.00810}, (2017).



  \bibitem{Alemi17}
   A.~A. Alemi, I.~Fischer, J.~V.~Dillon, and K.~Murphy, Deep variational
    information bottleneck, in {\em Proceedings of the International Conference
    on Learning Representations}, (Openreview, 2017).

  \bibitem{Hoffman2016}
  M.~D. Hoffman, M.~J. Johnson, and G.~Brain, {ELBO surgery: yet another way to
  carve up the variational evidence lower bound}, in {\em Workshop in
  Advances in Approximate Bayesian Inference, NIPS}, 2016.

  \bibitem{Onsager1944}
  L.~Onsager, Crystal statistics. i. a two-dimensional model with an
    order-disorder transition, Phys. Rev. {\bf  65}, 117 (1944).
  
  \bibitem{TensorFlow16}
  M.~Abadi, P.~Barham, J.~Chen, Z.~Chen, A.~Davis, J.~Dean, M.~Devin,
    S.~Ghemawat, G.~Irving, M.~Isard, M.~Kudlur, J.~Levenberg, R.~Monga,
    S.~Moore, D.~G. Murray, B.~Steiner, P.~Tucker, V.~Vasudevan, P.~Warden,
    M.~Wicke, Y.~Yu, and X.~Zheng, Tensorflow: A system for large-scale machine
    learning, in {\em 12th {USENIX} Symposium on Operating Systems Design and
    Implementation}, (USENIX Association, 2016).
  
  \bibitem{Kingma15}
  D.~P. Kingma and J.~Ba, Adam: A method for stochastic optimization, in {\em
  Proceedings of the International Conference on Learning Representations}, (Openreview, 2015). 
  
  \bibitem{GameTheory}
  J.~González-Díaz, I.~García-Jurado, and G.~Fiestras-Janeiro, \textit{An Introductory Course on Mathematical Game Theory}, (American Math. Society,  2010).

\bibitem {Assaad2008} F. Assaad and H. Everts, in \textit {Computational Many-Particle Physics, Lecture Notes in Physics}  (Springer, 2008), p. 277.

 \bibitem{Wei2017}
  L. Wei and Y.-F. Yang, Doping-induced perturbation and percolation in the two-dimensional Anderson lattice, {\textrm{Sci. Rep.}} \textbf{7}, 46089 (2017).
 
 \bibitem{Hu2019}
  D. Hu, J.-J. Dong, and Y.-F. Yang, Hybridization fluctuations in the half-filled periodic Anderson model, {\textrm{Phys. Rev. B}} \textbf{100}, 195133 (2019).

  \bibitem{Vaswani17}
  A.~Vaswani, N.~Shazeer, N.~Parmar, J.~Uszkoreit, L.~Jones, A.~N. Gomez, \L{}.
    Kaiser, and I.~Polosukhin, Attention is all you need, in {\em Advances in
    Neural Information Processing Systems}, (Curran Associates, Inc., 2017).
  
  \bibitem{parikh16}
  A.~Parikh, O.~T{\"a}ckstr{\"o}m, D.~Das, and J.~Uszkoreit, A decomposable
    attention model for natural language inference, in {\em Proceedings of the
    2016 Conference on Empirical Methods in Natural Language Processing}, (Association for Computational Linguistics, 2016).
  
  \bibitem{cheng16}
  J.~Cheng, L.~Dong, and M.~Lapata, Long short-term memory-networks for machine reading, in {\em Proceedings of the 2016 Conference on Empirical Methods in Natural Language Processing}, (Association for Computational Linguistics, 2016). 

  \bibitem{ioffe15}
  S.~Ioffe and C.~Szegedy, Batch normalization: Accelerating deep network
    training by reducing internal covariate shift, in {\em Proceedings of the
    32nd International Conference on Machine Learning}, (PMLR, 2015).

  
  \end{thebibliography}
\end{document}